\documentclass[12pt]{article} 
\usepackage[dvips]{graphics}
\title{Neutrino Proper Time?}  
\author{{\it Richard Shurtleff~}\thanks{affiliation and mailing 
address: Department of Mathematics and Applied Sciences, 
Wentworth Institute of Technology, 550 Huntington Avenue, 
Boston, MA, USA, ZIP 02115, telephone number: (617) 989-4338, fax 
number: (617) 989-4591 , e-mail address: shurtleffr@wit.edu}} 
\date{March 8, 2000}
\begin{document} 
          
\maketitle               

\begin{abstract}  

An electron neutrino can have the quantum phase of an electron, i.e. share its internal clock, if the neutrino takes a path in space-time that is not in the direction of its energy-momentum. Each flavor neutrino would then have a different internal clock; a muon neutrino would have a muon clock and a tau neutrino would have a tau clock. Perhaps surprisingly, there is some evidence suggesting neutrinos have  such clocks. If muon neutrinos travel on space-like paths then some atmospheric muon neutrinos would take such paths backwards into outer space and not be observed. These are lost at the source and have nothing to do with oscillations or flavor-changing in flight. The expected depletion of source muon neutrinos is shown here to be 9\%, which accounts for half of the missing muon neutrino source flux reported by Super-Kamiokande. Since there is no depletion in the electron neutrino flux source reported at SK and SN1987A electron neutrinos seem to have traveled at the speed of light, the electron neutrino travels on a light-like path. Accelerator-based experiments could be arranged to confirm the reverse motion of muon neutrinos.

 
			\end{abstract}
\pagebreak

\section{Introduction} \label{intro} 

	The quantum phase of a free electron on a given path $\Gamma$ is the product of its mass and the proper time along $\Gamma,$ divided by the constant $\hbar.$ Expressed in a general reference frame the product of mass and proper time, i.e. the action, is the integral of the space-time scalar product of energy-momentum and displacement. The classical path is a path that has a quantum phase that is stationary against small fluctuations $\delta \Gamma$ in the path. [1]

	If the path for a neutrino is light-like and in the direction of its light-like energy-momentum, then its quantum phase doesn't change, because the scalar product of its energy-momentum and displacement is then zero. This is the problem; the electron quantum phase on the electron's classical path is nonzero and proportional to its proper time while the neutrino quantum phase is completely different: the neutrino clock is frozen - it doesn't change as the neutrino moves.

	Hence, in Sec.~2 and 3, we consider neutrino paths with varying proper time, i.e. clocks that actually run. The neutrino proper time increases along the paths just as the proper time increases as an electron moves on its classical path. The neutrino quantum phase is then defined to be the proper time times the electron mass, just as it is for an electron. This definition makes the quantum phase of electrons and electron neutrinos agree. Electron, muon, and tau neutrinos have phases that disagree because the proper time is multiplied by the electron, muon, or tau mass, respectively. 

	In Sec.~4 we consider the possibility that muon neutrinos travel along space-like paths. We show that there is a loss of 7\% of the muon neutrinos from pion decays and a 10\% loss of muon neutrinos from muon decays, giving an average loss of 9\%. These 9\% carry their forward momentum backwards into space and are not detected. 

	Evidence for such an effect may have been found by the Super-Kamiokande collaboration (SK).[2] SK finds a depletion of atmospheric muon neutrinos compared with the expected flux. Determining the expected flux is in itself a complex problem.[3] At SK both the dependence on distance from source to detector and the overall normalization are measured. The observed overall normalization is 15.8\% with an estimated uncertainty of 25\%. Therefore the 9\% depletion derived in Sec.~4 is about half of the central value reported, 15.8\%. Thus the depletion of muon neutrinos at the source could be partly caused by some muon neutrinos moving in directions that differ from their momenta.

	No electron neutrino depletion is reported by Super-Kamiokande. In Sec.~5 we show that no depletion implies that the electron neutrinos have time-like or light-like paths. The arrival of electron neutrinos only a matter of hours ahead of light from SN1987A [4,5] is generally taken to mean that the neutrinos traveled to Earth from the supernova at light-speed or at near-light speed. Thus the light-like option for electron neutrinos explains both the SK observations and the SN1987A reports. 

	The predicted motions can be tested with current accelerator and detector technology. Arranging for a neutrino detector to be placed upstream from a pion or muon beam should detect the reverse motion muon neutrinos produced in the pion or muon decays, if such reverse motion muon neutrinos exist. 

	
\section{Neutrino Paths with Proper Time} \label{paths} 

	Before dealing with the neutrino, we discuss the path of an electron in space-time and how electron paths are related to quantum phase. Let $p^{\mu}$ be energy-momentum, $p^{\mu}$ = $\{p^{k},E\},$ and let $x^{\mu}$ be position, $x^{\mu}$ = $\{x^{k},t\},$ where $\mu \in$ $\{1,2,3,4\}$ and $k \in$ $\{1,2,3\}.$ Consider a path $\Gamma$ = $x^{\mu} (\tau),$ where $\tau$ is the proper time along $\Gamma,$ $(d\tau)^2$ = $(dt)^2$ $- dx^{k} dx^{k}.$ The quantum phase $\phi$ along $\Gamma$ is just $-S_{\Gamma}/\hbar.$ The classical path is a straight line that has a quantum phase of
\begin{equation}	\label{phi}
\phi = -\frac{S}{\hbar} = \frac{-Et + p^{k}x^{k}}{\hbar},  \hspace{1cm} {\mathrm{(electron)}} \hspace{1cm}
\end{equation}
 where $S$ is the action along the classical path and repeated indices are summed.

	The mass $m,$ $m >$ 0, determines the square of the energy-momentum, 
\begin{equation}	\label{mass}
  E^2 - p^{k} p^{k} = m^2 \hspace{1cm} {\mathrm{(electron)}} \hspace{1cm}
\end{equation}

	The action $S$ is invariant under space-time transformations of the 4-vectors $p^{\mu}$ and  $x^{\mu}$. So we can transform to a reference frame in which the momentum $p^{k}$ vanishes. We assume that $x^{k}$ is proportional to $p^{k}$ so that $x^{k}$ also vanishes. By (\ref{mass}), $E$ = $m.$ By (\ref{phi}), $t$ = $S/m,$ and, since this is the rest frame, the proper time $\tau$ is $S/m.$ Hence the square of the magnitude of $x^{\mu}$ in that frame as in all frames is 
\begin{equation}	\label{xtmag}
\tau^2 = t^2 - {x^{k}}^2 = \frac{S^2}{m^2} \hspace{1cm} {\mathrm{(electron)}} \hspace{1cm}
\end{equation}
By (\ref{phi}) and (\ref{xtmag}) the action on the classical path in the rest frame is $S$ = $Et$ = $m \tau,$ the electron mass times the proper time. In a general inertial frame we get
\begin{equation}	\label{Eptau2}
S = Et - p^{k} x^{k} = m \tau \hspace{1cm} {\mathrm{(electron)}} \hspace{1cm}
\end{equation}

	Assume that the electron and the electron neutrino have the same form for the quantum phase along the classical path,
\begin{equation}	\label{phinu}
\phi = -\frac{S}{\hbar} = \frac{-pt + p^{k}x^{k}}{\hbar},  \hspace{1cm} {\mathrm{(neutrino)}} \hspace{1cm}
\end{equation}
where the energy of the neutrino is the magnitude $p$ of its momentum, so that the energy-momentum is a null 4-vector.

	In this paper the electron and the electron neutrino share the same quantum phase, $m \tau.$ We get
\begin{equation}	\label{pptau2}
S = pt - p^{k} x^{k} = m \tau, \hspace{1cm} {\mathrm{(neutrino)}} \hspace{1cm}
\end{equation}
where $m$ is the electron mass. Each flavor has its own mass; for the muon or tau neutrino $m$ is the muon or tau mass.

		The alternative classical paths possible for a neutrino have sub- or super-luminal speeds or have the speed of light but in the direction opposite to the momentum. For space-like paths proper distance is more appropriate than proper time; I continue to say `proper time' even when I mean proper distance.

	We define $\tau$ to be the proper time as usual for time-like paths. For space-like paths it is actually the proper distance. We have 
\begin{equation}	\label{xtmagneu}
t^2 - {x^{k}}^2 = \eta \tau^2 , \hspace{1cm} {\mathrm{(neutrino)}} \hspace{1cm}
\end{equation}
where $\eta$ = $+1$ indicates a time-like path, $\eta$ = $-1$ indicates a space-like path, and $\eta$ = $0$ indicates a light-like path.  

	By  (\ref{xtmagneu}), a light-like path, $\eta$ = 0, implies that $x^{k}$ = $ \pm t n^{k},$ if $x^{k}$ and $p^{k}$ are  both in the direction of the unit 3-vector $n^{k}.$ For the solution $x^{k}$ = $ + t n^{k},$ and by (\ref{pptau2}), the proper time $\tau$ is stuck at zero. In this paper the underlying presumption is that the action for an electron and its neutrino are one and the same function of proper time. Hence we dismiss the possibility that a neutrino path can have the solution $x^{k}$ = $ + t n^{k},$ because no electron path has zero proper time. 

	It may be that the neutrino follows a space-like path, i.e. $\eta$ = $-1$ in (\ref{xtmagneu}). In this case the electron and its neutrino do not travel the same kind of path; the electron's classical path is time-like, $\eta$ = $+1.$  This is true for the classical paths. But in quantum mechanics we can consider paths $\Gamma$ that are space-like even for electrons. Thus having the electron and its neutrino share the same internal clock implies extending the formulas to space-like neutrino paths.

\section{Neutrino Paths in Two Frames} \label{paths2} 

	Relativity protects scalar products and 4-vector magnitudes, but directions in space are not preserved under Lorentz transformations. If neutrino momentum and displacement can be in different directions, then only in special frames do momentum and displacement align, dubbed `Aligned Momentum' (AM) frames. `Aligned' is used in the following ways: the momentum is `in the same direction' as the displacement or the momentum is `in the opposite direction' of the displacement. A boost in the direction of the momentum and displacement takes one AM frame to another AM frame.

\begin{figure}[h] \label{AM}	
\vspace{0in}
\hspace{0in}\includegraphics[0,0][288,178]{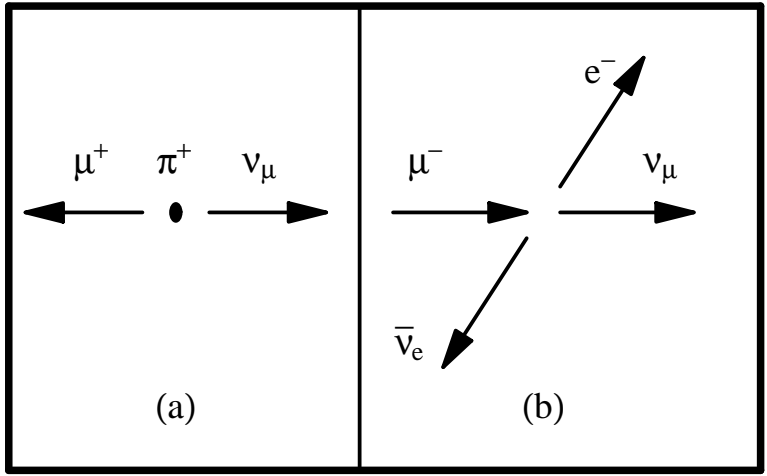}
\caption{ In an AM frame the neutrino displacement is parallel or antiparallel to its momentum. One Aligned Momentum reference frame is assumed to be the frame with muon momentum equal to neutrino momentum. (a) In charged pion decay the muon and muon neutrino have momenta that are equal and opposite in the pion rest frame. (b) In muon decay the momenta of the muon and muon neutrino can be equal and in the same direction. This AM frame for the muon neutrino is also an AM frame for the electron neutrino.}
\end{figure}

	In an AM frame we introduce new notation $X$ and $T$ defined by
\begin{equation}	\label{XTdef}
x^{k} = X \frac{ \tau }{2} n^{k} \hspace{1cm} t = T \frac{ \tau}{2}.\hspace{1cm} {\mathrm{(AM)}} \hspace{1cm}
\end{equation}
where $n^{k}$ is a unit vector in the direction of the momentum, $p^{k}$ = $p n^{k}.$ By (\ref{pptau2}) and (\ref{xtmagneu}) we have
  \begin{equation}	\label{xt}
 X = \eta s - \frac{1}{s} \hspace{1cm} T =  \eta s + \frac{1}{s} \hspace{1cm} {\mathrm{(AM)}} \hspace{1cm}
\end{equation}
where $s$ stands for the ratio of momentum to mass, $s$ = $p/m$ (= $\sinh{w},$ where $\tanh w$ is the speed of a particle of mass $m$ and momentum $p.$) In Fig.~2 the path $(X,T)$ is plotted for the three values of $\eta.$

\begin{figure}[h] \label{etasplot}	
\vspace{0in}
\hspace{0in}\includegraphics[0,0][288,180]{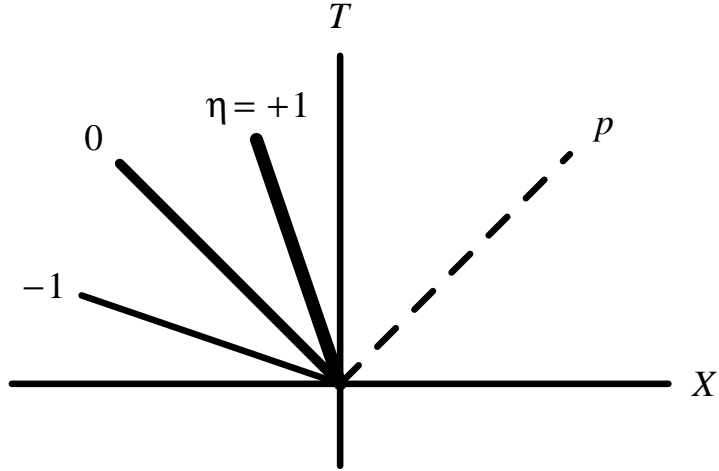}
\caption{Paths for $s$ = $p/m$ = 0.7 in the AM frame, i.e. with momentum and displacement aligned, both in the $\pm X$-direction. For a time-like ($\eta$ = +1), light-like ($\eta$ = 0), or space-like ($\eta$ = $-1$) path, the neutrino moves in the direction opposite to the momentum $p^{k}$ that it carries. ($p^{\mu}$ is a light-like 4-vector in the direction of the dashed line.) The displacements satisfy the requirement, (\ref{pptau2}), that they have a nonzero space-time scalar product  with the energy-momentum.}
\end{figure}

	The AM frame and a general frame, the `LAB' frame, differ by a space-time transformation, i.e. a rotation followed by a boost. Choose the positive $x$-axis as the direction of the boost. Let $\alpha $ be the angle between the direction $X,$ i.e. the direction $n^{k}$ of the momentum in the AM frame, and the positive $x$-axis. See Fig.~3a. We get
   \begin{equation}	\label{xmuprime}
X^{\prime} = \gamma [ X \cos\alpha +  \beta T ] \hspace{0.5
cm} Y^{\prime} = X \sin\alpha \hspace{0.5cm} T^{\prime} = \gamma [ T  +  \beta X \cos\alpha] \hspace{0.5cm} {\mathrm{(LAB)}} \hspace{0.5cm}
\end{equation}
where a prime indicates a value in the LAB frame, $\beta$ is the boost velocity, and $\gamma$ = $(1 - \beta^2)^{-1/2}.$
The neutrino energy-momentum in the LAB frame is
   \begin{equation}	\label{pmuprime}
p_{x}^{\prime} = \gamma [ p \cos\alpha +  \beta p ] \hspace{1cm} p_{y}^{\prime} = p \sin\alpha \hspace{1cm} p^{\prime} = \gamma [ p  +  \beta p \cos\alpha] \hspace{1cm} {\mathrm{(LAB)}} \hspace{1cm}
\end{equation}

\begin{figure}[h] \label{CMLAB}	
\vspace{0in}
\hspace{0in}\includegraphics[0,0][288,178]{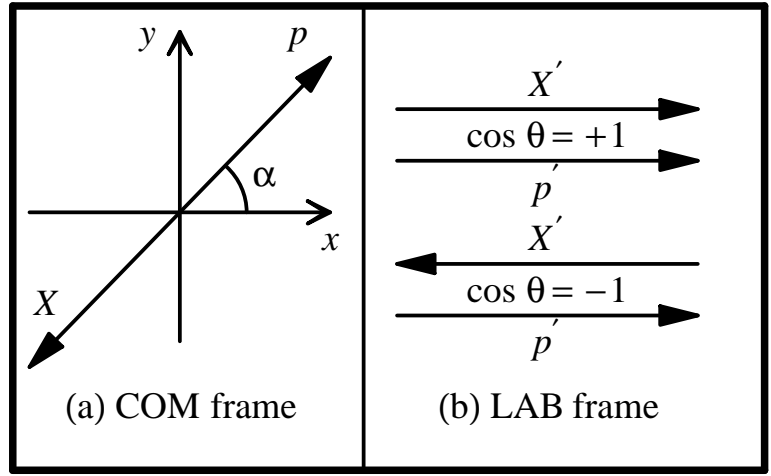}
\caption{The AM frame is boosted in the $x$-direction to the LAB frame. (a) Let $\alpha$ be the angle between the $x$-direction and the momentum $p.$ (b) For neutrinos with much larger momenta in the LAB frame than in the AM frame, the boosted momentum and displacement are either parallel or opposite, i.e. $\cos \theta$ is either $+1$ or $-1,$ for the most part. See Fig.~5 and 6.}
\end{figure}

\section{Muon Neutrinos} \label{muonneu} 

	Cosmic rays continually bombard the Earth's atmosphere creating muon and electron neutrinos as well as other particles. A muon neutrino is produced together with a muon in charged pion decay and together with an electron and an electron neutrino when a muon decays. Other neutrino-producing processes are said to occur much less often. 

	In the Super-Kamiokande (SK) experiment atmospheric neutrinos are detected, muon neutrinos from pion and muon decays and electron neutrinos from the muon decays. We consider muon neutrinos first. We consider only what happens to muon neutrinos just after they are created at the source. This discussion has no bearing on the depletion of muon neutrinos in flight from the source in the atmosphere to the SK detector.

	In the rest frame of the decaying particle, all directions in space are equivalent. Hence the rest frame, i.e. the Center of Mass (COM) frame, is convenient for finding average momenta. See Fig.~4.

	In the decay of a pion or muon, the muon and the muon neutrino can share a common momentum and be aligned only in one special reference frame. To make the muon and muon neutrino even more alike in this frame assume that this special frame is also an Aligned Momentum (AM) frame so that the muon neutrino and its displacement are aligned. The muon momentum and the muon neutrino momentum are equal in magnitude and aligned and the muon and muon neutrino displacements are also aligned. See Fig.~1.

\begin{figure}[h] \label{CM}	
\vspace{0in}
\hspace{0in}\includegraphics[0,0][288,178]{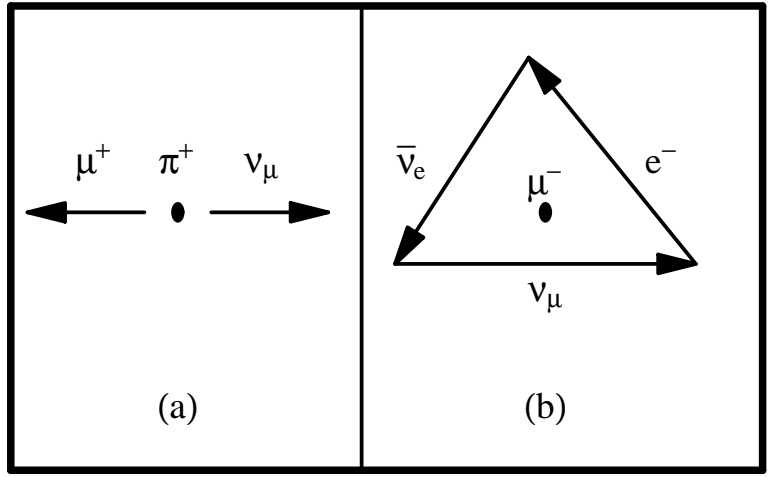}
\caption{The Center of Mass reference frame. The arrows indicate momenta, not displacements. (a) In charged pion decay the muon and muon neutrino have momenta that are equal and opposite. (b) In muon decay the momenta of the muon neutrino, electron, and electron neutrino combine to form a triangle with a constant perimeter when the electron mass is neglected. }
\end{figure}

\begin{figure}[h] \label{px1}	
\vspace{0in}
\hspace{0in}\includegraphics[0,0][288,178]{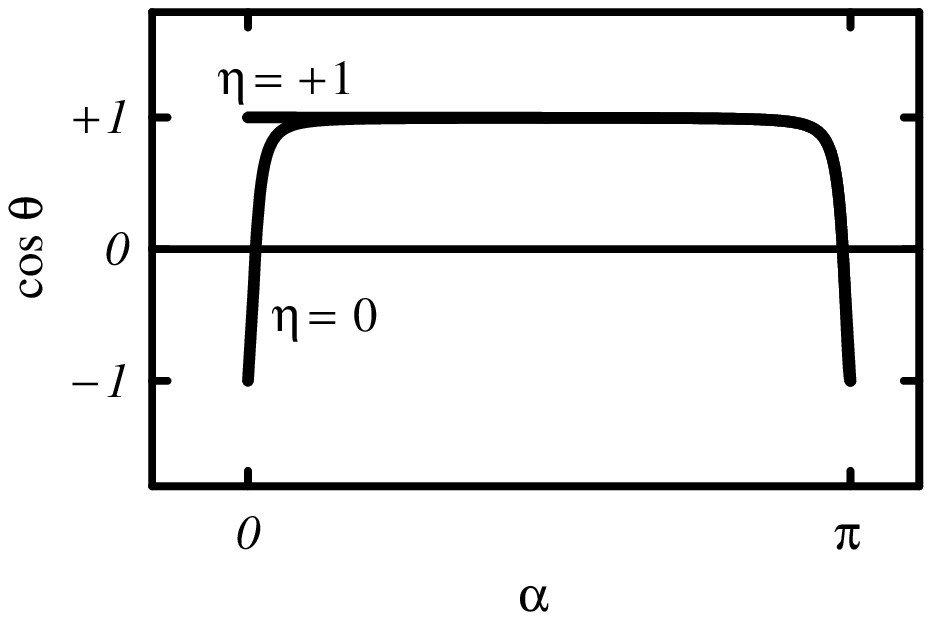}
\caption{Motion is mostly parallel to momentum in the LAB frame for time-like and light-like paths. The angle $\alpha$ is the angle between the neutrino momentum $p$ and the direction of the boost in the AM frame, see Fig.~3a. $\theta$ is the angle between the momentum $p^{\prime}$ in the LAB frame and the displacement $X^{\prime}$ in the LAB frame. In this sketch, with a $\gamma$ of 25, neutrinos traveling along light-like or time-like paths mostly move forward in the LAB frame, i.e. $\cos \theta$ = 1 and $\theta$ = 0. }
\end{figure}

	For muon neutrinos originating in pion decay, the pion rest frame is both the Center of Mass frame in which neutrinos are emitted in all directions equally and the AM frame in which the neutrino momentum and muon momentum are equal in magnitude. Let the SK rest frame be the LAB frame. Since the pion rest frame and the LAB frame typically differ by a large boost, the direction of the boost from the AM to the LAB frame, the $x$-direction in Fig.~3a, is along the line from the pion to SK. And, in the pion rest frame, the angle $\alpha $ is the angle between the emitted neutrino and the $x$-direction toward SK.

	For muon neutrinos detected at SK with momenta of 400 MeV/$c$ or more, the factor $\gamma$ is more than 10. By (\ref{xmuprime}) and (\ref{pmuprime}) and as shown in Fig.~5 and Fig.~6, for large factors $\gamma >>$ 1, the displacement ${X^{\prime}}^{k}$ is mainly either in the same direction as or in the direction opposite to the momentum ${p^{\prime}}^{k}$ in the LAB frame.  

\begin{figure}[h] \label{px2}	
\vspace{0in}
\hspace{0in}\includegraphics[0,0][288,178]{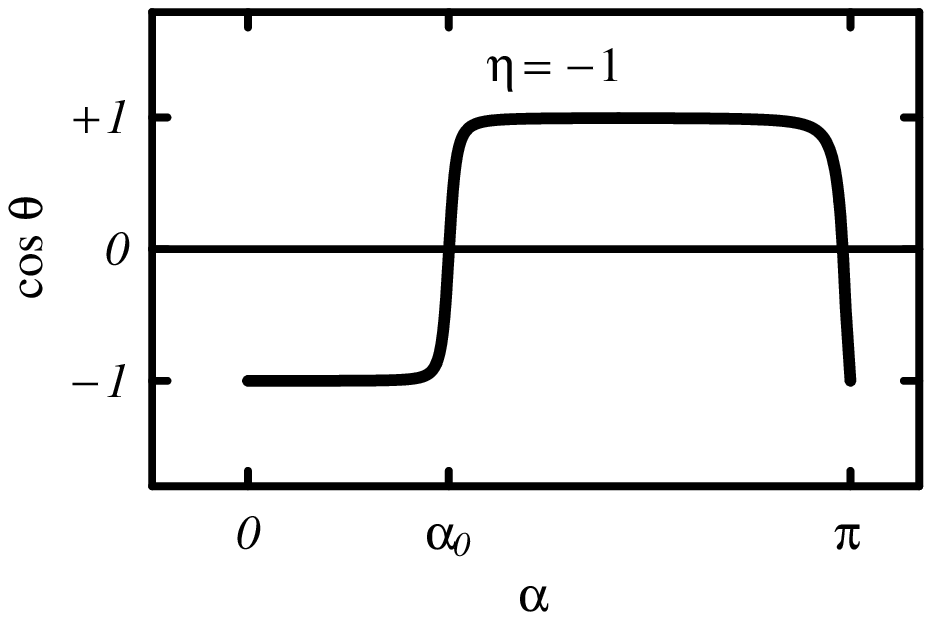}
\caption{Motion is opposite to the momentum in the LAB frame for space-like paths with direction $\alpha <$ $\alpha_{0}.$ For such paths, $\cos \theta$ = $-1$ and $\theta$ = $180^{{\mathrm{o}}}.$ Atmospheric neutrinos with  $\alpha <$ $\alpha_{0}$ and momentum pointed towards the Earth in the LAB (Earth) frame would move backwards into outer space. }
\end{figure}

	From Fig.~6, we see that reverse motion, i.e. ${x^{\prime}}^{k}$ opposite to momentum ${p^{\prime}}^{k},$ occurs for space-like paths, $\eta $ = $-1.$ By (\ref{xt}), (\ref{xmuprime}), and (\ref{pmuprime}) we find that $X^{\prime}$ is opposite to ${p^{\prime}}^{k}$, i.e. $\cos \theta$ = $-1,$ for directions with $\alpha < \alpha_{0}$ where 
   \begin{equation}	\label{alpha0}
\cos \alpha_{0} =  -  \frac{ T}{X} =   \frac{1-s^2}{1+s^2}.  \hspace{1cm} (\eta = -1) \hspace{1cm} (\gamma >> 1) \hspace{1cm}
\end{equation}
Therefore muon neutrinos emitted in the direction of SK out to an angle of $\alpha_{0}$ actually move back toward outer space in the SK frame of reference. These muon neutrinos cannot be detected at SK.

	Assume that muon neutrinos travel along space-like paths, i.e. choose $\eta$ = $-1$ in (\ref{xtmagneu}). Since all directions $\alpha$ in the pion frame are equally likely, we calculate the expected normalization $N$ to be the fraction of the unit sphere for which the path of the muon neutrinos is toward SK. We get
   \begin{equation}	\label{normal}
N = \frac{{\mathrm{toward}}\:\:{\mathrm{SK}}}{{\mathrm{all}}} = \frac{\int_{\alpha_{0}}^{\pi} 2 \pi \sin \alpha d \alpha}{4 \pi} =  \frac{1}{1+s^2}. \hspace{1cm} (\eta = -1) \hspace{1cm}
\end{equation}

	In the rest frame of a decaying pion, by conservation of energy and momentum and the known rest masses of the pion and muon, we find that the ratio of muon neutrino momentum to muon mass is $s$ = $p/m_{\mu}$ = $(m_{\pi}^2 - m_{\mu}^2 )/(2 m_{\mu} m_{\pi})$ = 0.28. By (\ref{normal}) the normalization $N$ for muon neutrinos from pion decay is $N$ = 0.93. 

	Atmospheric muon neutrinos are also produced in muon decay. A muon decays to an electron, an electron neutrino, and a muon neutrino. The electron mass is only 1/210th of the muon mass, so let us neglect the electron mass. Then we can treat the three decay particles equally, since each decay particle has a null energy-momentum 4-vector. The energy of each decay particle is the magnitude of its momentum. In the rest frame of the muon, the three momenta cancel. Thus the three momenta form a triangle with a constant perimeter equal to the muon mass, see Fig.~4b. An equilateral triangle is the symmetric solution with no one decay particle being given any more or less momentum than any other decay particle. Thus an estimate of the average muon neutrino momentum is about one-third of a muon mass, $p_{\nu}$ = $m_{\mu}/3.$  

	But the muon rest frame is also an Aligned Momentum frame for the muon neutrino. This follows because, as we already assumed, the muon neutrino is certain to have its momentum and displacement aligned in the frame in which the muon and the muon neutrino have aligned momenta with equal magnitudes as in Fig.~1b.  Furthermore the transformation from the AM frame with equal momenta, i.e. Fig.~1b, to the muon rest frame, Fig.~4b, is a boost that is itself aligned with the momenta and displacements. Hence the muon rest frame is an Aligned Momentum frame for the muon neutrino. 

	Since the muon rest frame is an AM frame and since all directions in the rest frame are equivalent, we can use (\ref{normal}) to calculate the normalization. The estimate above gives $s$ = $p/m_{\mu}$ = $1/3.$ By (\ref{normal}) we get $N$ = $0.90$. 

	The overall normalizations estimated for muon neutrinos from pion and muon decay are collected in the following table.
 
\vspace{0.3cm}
\begin{center}\begin{tabular}{|c|c|c|} \hline
$\nu_{\mu}$ Source & $s$ = $p/m_{\mu}$ & $\nu_{\mu}$ Normalization\\[0.5ex]
\hline\hline
$\pi^{-} \rightarrow$ $\mu^{-} + {\bar{\nu}}_{\mu}$ & $0.28$ & 0.93\\[0.5ex]
\hline
$\mu^{-} \rightarrow$ $e^{-} + {\bar{\nu}}_{e} + \nu_{\mu}$ & $0.33$ & 0.90\\
\hline\end{tabular}\end{center}
\vspace{0.3cm}

	Together the pion and muon sources of muon neutrinos give an average normalization of
   \begin{equation}	\label{normal2}
\bar{N} = \frac{0.93 + 0.90}{2} = 0.91, \hspace{1cm} {\mathrm{(neutrino)}} \hspace{1cm} (\eta = -1) \hspace{1cm}
\end{equation}
which is an 9\% depletion. The best fit value of the source depletion at Super-Kamiokande is an overall normalization of 15.8\% with an estimated 25\% uncertainty.[2] Thus the observed Super-Kamiokande depletion can be partly explained by muon neutrinos that move along space-like paths carrying light-like momenta. If future observations bring the 15.8\% depletion down closer to 9\%, then some other muon depletion mechanism would not be needed.

	An accelerator based experiment testing the model could be devised. By the discussion of pion and muon decays above, arranging for a detector to be placed upstream of a pion or muon beam should detect upstream muon neutrinos. These muon neutrinos carry light-like momenta backwards, opposite to the direction of the neutrino momentum. Thus experiments should look for neutrino-induced events upstream that deposit downstream momenta.

\section{Electron and Tau Neutrinos} \label{electronneu} 

	Atmospheric electron neutrinos are produced in muon decays. By the discussion in Sec.~\ref{muonneu} the electron neutrino is emitted in all directions equally likely in the muon rest frame, the COM frame. And the average electron neutrino has a momentum of $m_{\mu}/3$ the same as the muon neutrino. For an electron neutrino the momentum to mass ratio is very large, $s$ = $p/m_{e}$ = $m_{\mu}/(3 m_{e})$ = 70. By (\ref{normal}), if the electron neutrinos move on space-like paths, then 99.98\% of atmospheric electron neutrinos move in the direction opposite to their momenta in the SK LAB frame. These would not be seen at SK. 

	But no electron depletion is reported by SK. Hence electron neutrinos cannot move on space-like paths, $\eta \neq$ $-1.$ From Fig.~5 we see that time-like and light-like paths are still open to the electron neutrino since virtually all neutrinos with $\eta$ = $+1$ or $\eta$ = 0 would move in the direction of their momenta and arrive at the SK detector. 

	The detection of neutrinos just hours before photons from the supernova SN1987A  is strong evidence that neutrinos traveled on light-like or at least near light-like paths in the direction from SN1987A toward Earth. Furthermore the momentum absorbed in the detectors pointed away from SN1987A. The displacement and energy-momentum appear to be null 4-vectors pointing in the same direction. By Fig.~5 these electron neutrinos could have $\eta$ = $0$ if the factor $\gamma$ for the boost from the COM frame to the LAB frame is large enough. We assume that it is and conclude that electron neutrinos move along light-like paths. 

	Tau neutrinos have not been detected, as far as I know. By the discussion of muon decay, a tau neutrino would have an average momentum of $p$ = $m_{\tau}/3$ = $592$ MeV in the tau rest frame. For taus created essentially at rest in the LAB frame, the tau neutrinos would follow one of the paths in Fig.~2 but with $s$ = $0.33.$ Hence, for a time-like path, a space-like path, or a light-like path, all the tau neutrinos would travel in the direction opposite to the momentum carried by the tau neutrino. Experiments verifying this behavior would find tau neutrino induced events with momenta pointing towards the tau source, not away from the source.

\appendix

 \section{Problems} 

\noindent 1. Consider muon decay in the case when the muon neutrino has the average muon neutrino momentum of $p$ = $m_{\mu}/3$ in the muon rest frame, as found in Sec.~4. Find the relative velocity of the AM frame with equal momenta, Fig.~1b, with respect to the muon rest frame, Fig.~4b.

\vspace{0.3cm}

\noindent 2. Show that there is no AM frame with equal momenta for neutrino scattering, i.e. no frame such as the ones illustrated in Fig.~1. The momentum before the interaction cannot be aligned with the momentum after the interaction and keep the same magnitude.

\end{document}